**Nate Schnitzer**
LEMMA,
University of Maryland,
College Park, Maryland,
nschnitz@terpmail.umd.edu

**Ethan Ewing**
LEMMA,
University of Maryland,
College Park, Maryland
eewing1@terpmail.umd.edu

**Hung Anh Vu**
LEMMA,
University of Maryland,
College Park, Maryland,
hvu1@terpmail.umd.edu

# Surjective One Dimensional Span 6 Cellular Automata

*Using FSA and the construction algorithm, we generated a list of surjective span 6 cellular automata as a modest sample for our FDense program. We wanted to experimentally quantify Mike Boyle's conjecture which states that the jointly periodic points of a one dimensional cellular automata are dense. Futhermore, we wanted to know if the cardinality of a cellular automata on N symbols is $\geq \sqrt{N}$.*

## 1 Introduction

In this paper we are exploring one dimensional cellular automaton (CA) through our own computer programs to try and get a better understanding of their periodic orbits. The main focus is on surjective one dimensional CA highlighted in [1]. Mike Boyle and Bryant Lee developed a computer program for studying one-dimensional cellular automata to gather experimental data to justify conjecture 1.1 [1]. Conjecture 1.1 is a very well-known open question because there seems to be a correlation between the behavior of cellular automata and dynamical systems.

**Conjecture 1.1** For every surjective one-dimensional cellular automaton, the jointly periodic points are dense.

Through our research, we wish to replicate the results from [1] and also provide more evidence for conjecture 1.1 by experimenting on span 6 surjective maps. Additionally, we wish to know if the fraction of surjective one-dimensional cellular automata given by range of n codes, which have dense jointly periodic points, goes to 1 as n goes to infinity.

If we let $P$ represent the number of jointly periodic points in $Per(f) \cap P_k(SN)$ (i.e. $P = |Per(f|Pk(SN))|$ ) and set $V_k(f, SN) = P^{1/k}$, we can define

$$V(f, S_N) = \lim_k \sup V_k(f, S_N)$$

This raise the question if every cellular automata ($f$) defined on N symbols $V(f, S_n) \geq \sqrt{N}$ or if $V(f, S_n) \geq 1$.

## 2 Definitions

Cellular automata (CA) are mathematical models that exhibit complex behavior through the interaction of simple units, known as cells, which are arranged in a regular grid. Cellular automata have several key properties that make them interesting and useful in various areas of science and engineering. Some of these properties are:

(1) Locality: Each cell in a CA only depends on its local neighborhood of adjacent cells. This locality property makes CAs computationally efficient and easy to parallelize, as each cell can be updated independently based on its local state.

(2) Homogeneity: The same local rule is applied to every cell in the CA. This property allows the CA to exhibit complex global behavior that arises from the repeated application of simple local rules.

(3) Emergence: The global behavior of a CA emerges from the interaction of many simple local rules. This can lead to the emergence of complex patterns, structures, and behaviors that are not immediately obvious from the local rules.

**Definition:** Let $\Sigma_N$ denote the set of doubly infinite sequences $x = ...x_{-1}x_0x_1...$ such that each $x_i$ lies in a finite alphabet $A$ of $N$ symbols. A one-dimensional cellular automata (CA) is a mapping $f : \Sigma_N \to \Sigma_N$ which is defined by a local rule $F : A^{2M+1} \to A$ where $M > 0$, such that for all i $(f(x))_i = F(x_{i-M}, ..., x_{i+M})$ [1]

**Definition:** Let $\sigma(x)_i$ denote the shift map on a sequence. $\sigma(x)_i = x_{i+1}$. Let $S_N$ denote the shift map on $\Sigma_N$ [1]

One single application of the shift map shifts every cell of $x \in \Sigma_N$ to the right one index. The shift map moves the symbols of a sequence without changing the content of the sequence itself.

We refer to the points of (not necessarily least) period k of a map S as $P_k(S)$, and the points fixed by $S^k$ as Per(S), where Per(S) = $\cup_k P_k(S)$. Thus, Per($S_N$) is the set of spatially periodic points for a one-dimensional cellular automaton on N symbols. The jointly periodic points of a cellular automaton map f on N symbols are the points in Per(S) that are also periodic under f, i.e., the points that are temporally periodic as well as spatially periodic.[1]

In simpler terms, an $x \in \Sigma_N$ is spatially periodic if $x = \sigma^n(x)$ i.e. the point is equal to the point shifted n times. An $x \in \Sigma_N$ is temporally periodic if $x = f^n(x)$. An $x \in \Sigma_N$ is jointly periodic if $\sigma^n(x) = x = f^m(x)$.

**2.1 Density.** One of our motivations is to quantify Mike Boyle's conjecture which states that the jointly periodic points of one-dimensional cellular automata are dense. In order to do this, we must have a firm grasp of what it means to be dense in $\Sigma_2$. Generally, when we talk about density of sets, a set $Y \subseteq X$ is dense in $X$ if every point in $X$ has a point in $Y$ that is arbitrarily close. For metric spaces, such as $\Sigma_2$, a distance function, $d(x, y)$ for $x, y$ in the metric space $(X, d)$, can be used to determine if points are arbitrarily close. If $d(x, y) < \epsilon$ for $\epsilon > 0$, then $x$ $y$ are arbitrarily close. For $\Sigma_2$,

$$d(x, y) = \sum_i^\infty 2^{-i} * |x_i - y_i|$$

Since $\Sigma_2$ is the set of doubly infinite sequences, to apply $d(x, y)$, it is necessary to choose a starting index.

**Example:** Prove the set spatially periodic points are dense in $\Sigma_2$.

Proof: Let $x$ be an arbitrary point in $\Sigma_2$. $x = x_1, x_2, x_3, ...$ Let $y = x_1, x_2, ..., x_{n-1}, x_1, x_2, ..., x_{n-1}, x_1, ....$ Clearly, y is spatially periodic. $d(x, y) = 2^{-n}$. Let $\epsilon = 2^{-n}$. Clearly $\epsilon > 0$ and approaches 0 as $n$ goes to infinity. Since, y is an arbitrary spatially periodic point and x is arbitrary in $\Sigma_2$, and $d(x, y) \leq \epsilon$, the set of spatially periodic points are dense in $\Sigma_2$.

A subset E of $\Sigma_N$ is considered dense if for every point x in $\Sigma_N$ and every $k \in N$ there exists y in E such that $x_i = y_i$ whenever $|i| \leq k$. We say E is m-dense if every word of length m on symbols from the alphabet occurs in a point of E. This definition of density is more closely linked to conjecture 1.1. This will be more clear and applicable when we discuss the FDense program.

**2.2 Surjectivity.** Since we are concerned with surjective cellular automata, we should have a working definition of what it means for a cellular automata is surjective. For a map $F : A \rightarrow B$ to be surjective, all elements of set $B$ must be attained by applying $F$ to some $a \in A$.

**2.3 Span and Tabular Representation.** We can define a cellular automata as a polynomial function such as $f = x_{-1} + x_0 x_1 (mod 2)$. This means that the next generation of $f$ is defined as $(fx)_i = 2x_{i-1} + x_i(x_{i+2})(mod 2)$. The span of this CA is 1 plus the maximum difference of coordinates with nonzero coefficients [1]. For our example, it is 1+ 2(1) = 4 so our CA is span 4. To make our implementation easier, we can represent the block code $(f)$ in tabular form instead of polynomial form. The tabular form of our function defines a look up table which allows for us to "look up" the output of any input. Below is a description of converting polynomial into tabular [1].

```
Suppose we want to specify the function f = x[0] + x[1]*x
    [2] as a table. First, note that the function is of
    span 3. Enumerate the values of the function for
    values of x[0], x[1], and x[2], going through them
    in lexicographic order as demonstrated below.
    Notice that incrementing starts at x[2] rather than
    x[0].

     0 1 2 3 4 5 6 7
x[0] 0 0 0 0 1 1 1 1
x[1] 0 0 1 1 0 0 1 1
x[2] 0 1 0 1 0 1 0 1
f    0 0 0 1 1 1 1 0

Then the tabular rule is "0001 1110".
```

## 3 Simple CA Examples

**3.1 Elementary CA.** Elementary cellular automata exist in $\Sigma_2$, the set of doubly infinite sequences x = $...x_{-1}x_0x_1...$ where each $x_n$ is a 0 or 1. For span 3, each window is 3 bits, thus there are $2^3 = 8$ different possibilities for each window. Each window will produce a 0 or 1. We can classify the map $f : \Sigma_2 \rightarrow \Sigma_2$ with an 8-bit binary number.

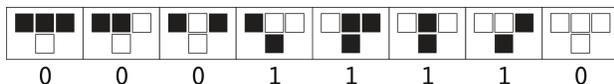

This map would correspond to the 8-bit binary number 00 011 110. 00 011 110 is 30 in base 10, so the Wolfram Code[3] for this mapping is rule 30. The mapping that turns all windows into black would be 11 111 111, rule 255. Likewise, the mapping that turns all windows into white 00 000 000, rule 0. Since there are 8-bits and $2^8 = 256$, there are 256 different rules.

**3.2 Python Implementation.** One dimensional Cellular Automata are very simple structures to implement in code using a sliding window approach, where an iterator moves through an array, performing some action based on a subset of the array determined by the current position.

Suppose you are given a CA Rule of width 3 and a 6 element long sequence, A, to apply the rule to. So, at a given index i, you would take the window/subset [A[i-1], A[i], A[i+1]] and apply the rule against that window. You would map this against the entire sequence, getting the newly evolved CA state.

Using python, we can represent the local function using a dictionary, where k-length tuples can be mapped to some list. We can generate the dictionary representing a given rule using the *get_rule_dict* function that we created:

```python
import itertools

def get_rule_dict(n: int, width: int):
    # Get the list representing the rule num
    # Converts n to binary string of length 2^3 and then
      pulls it back into
    # int list
    vals = [int(x) for x in '{0:0{k}b}'.format(n, k=2**
      width)]

    # Create a dictionary mapping the possible length
      width binary sequences to the rule and returns it
    return dict((zip(itertools.product([1,0], repeat=
      width), vals)))
```

Now that we have then dictionary containing the rule's mapping, we can create a function that applies this to some sequence. As said before, we want to use a sliding window technique to apply this mapping to the function on a window of width k.

We created a function *apply_function* which takes in a list of integers representing the sequence for the rule to be applied to, a dictionary containing the mapping of the binary sequences to the output in the form of tuples and ints, an int representing the width of the rule.

```python
def apply_mapping(line, rules, width):
    new_line = []
    for i in range(0, len(line) - (width - 1)):
        window = tuple(line[i:i+width])
        new_line.append(rules[window])
    return tuple(new_line)
```

Additionally, we can create another function to generate a random binary sequence of length n:

```python
def get_random_seq(n):
    seq = []
    for _ in range(n):
        seq.append(random.randint(0,1))
    return seq
```

Using these three functions, it is trivial to create an interface that can be used to represent Cellular Automata in Python.

For example, we can use this to run 10 Evolutions of Rule 30 on a random sequence of length 25:

```python
seq  = get_random_seq(25)
rule_dict = get_rule_dict(30,3)
print("Initial Sequence: {}".format(seq))
for i in range(10):
    seq = apply_mapping(seq, rule_dict, 3)
    print("Evolution {}: {}".format(i+1, seq))
```

## 4 Finite Automata Representation

Typically, we think of Cellular Automata and the local rules in terms of a mathematical function or mapping. However, it is sometimes more useful to view Cellular Automata rules in a more abstract manner. We can think of the local transition rule of the cellular automata as a function that takes in the current state of a cell and its neighboring cells and returns the next state of the central cell. Using this definition, it is easier view our local rules in terms of finite automata.

Finite Automata (FA) are a computational model which consists of a finite set of states, input states, final states, and a transition function which maps an input state to a new state. The computational model is typically represented as a directed graph. There are 2 classes of finite automata relevant to our work: Non-Deterministic Finite Automata (NFA) and Deterministic Finite Automata (DFA). DFAs are a subset of NFAs. DFAs have a special restriction where they can only have 1 transition per state and must end in a final state. Typically, FA are defined by a tuple $(\Sigma, Q, q_0, F, \delta)$:

- Alphabet, typically denoted by $\Sigma$

    - For example, a binary alphabet is $\Sigma = \{0, 1\}$

- Set of states, typically denoted by $Q$

- Starting state, typically denoted by $q_0$

    - $q_0 \in Q$

- Final States, denoted by $F$

    - $F \subset Q$

- The transition function, denoted by $\delta$

    - $\delta : Q \times \Sigma \to Q$

In his paper "Computation Theory of Cellular Automata", Stephen Wolfram discusses how to go from local mappings for cellular automata to NFAs and DFAs. He begins by demonstrating for rule 76 which has the following mappings:

$111 \to 0$, $110 \to 1$, $101 \to 0$, $100 \to 0$ $011 \to 1$, $010 \to 1$, $001 \to 0$, $000 \to 0$.

The first NFA here is the most basic build of an NFA for rule 76. Each node is a pair of possible values. Each transition turns this pair of values into a 3-tuple of values which we can then use the local mapping to get to the next state and generate part of the output. For example, lets say we want to know what f(01011) is where f is the mapping for rule 76. Since the first two values are 01, we begin in node 01. Our next value of our input is 0, so we take the 010 transition from the 01 node, which outputs a 1 and takes us to the 10 node. The next value of our input is 1 so we will take the 101 transition from 10, which outputs 0 and takes us to the 10 node. Continuing, we will now take the 011 transition from 01, which outputs a 1 and brings us to the 11 node. We are now at the end of our input string. By tracing the NFA, we know f(01011) = 101.

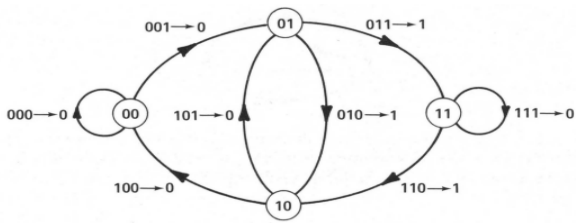

One thing to notice about this NFA is that the 00 and 01 nodes only have paths that output a 0 pass through them. Because of this, we are able to reduce our NFA to have less nodes. Here is the NFA with the 00 and 01 nodes combined:

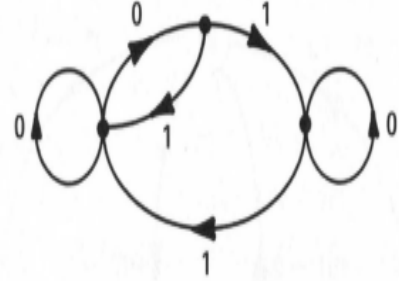

To get the DFA of rule 76, we begin from a start node 00, 01, 10, 11. From this node, we can get to 00, 01, 11 by a 0-arc transition. A 0-arc transition, is a transition that outputs a 0. For example, 00 to 00. From 00, 01, 10, 11, we can get to 10, 11 by a 1-arc transition. We repeat this process for all resulting sets until we get no new sets. ie. we will now do this for 00, 01, 11 and 10, 11. Below is Wolframs full construction of states and transitions for the DFA for rule 76. He defines the 00 node as $u_0$, the 01 node as $u_1$, the 10 node as $u_2$, and the 11 node as $u_3$.

$$\psi_S = \{u_0, u_1, u_2, u_3\} \to 0 \{u_0, u_1, u_3\}, \quad \{u_0, u_1, u_2, u_3\} \to 1 \{u_2, u_3\},$$
$$\{u_0, u_1, u_3\} \to 0 \{u_0, u_1, u_3\}, \quad \{u_0, u_1, u_3\} \to 1 \{u_2, u_3\},$$
$$\{u_2, u_3\} \to 0 \{u_0, u_1, u_3\}, \quad \{u_2, u_3\} \to 1 \{u_2\},$$
$$\{u_2\} \to 0 \{u_0, u_1\}, \quad \{u_2\} \to 1 \{\},$$
$$\{u_0, u_1\} \to 0 \{u_0, u_1\}, \quad \{u_0, u_1\} \to 1 \{u_2, u_3\}.$$

Notice that $u_2 \to 1 \emptyset$. This means from the $u_2$ state there will be no 1 transition. Here is the DFA for rule 76.

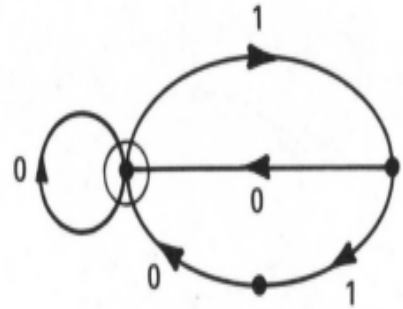

**4.1 The Construction Algorithm.** One paper that we inspected was "Decision Procedures for Surjectivity and Injectivity of Parallel Maps for Tessellation Structures" by S. Amoroso and Y. N. Patt. In this paper the authors build a decision procedure to determine surjectivity for one-dimension cellular automata. They call this procedure the Construction Algorithm.

For our purposes, we will begin by selecting an element $b \in \Sigma_2$ and let the node at level 0 be the set of all 3-tuples $a_1 a_2 a_3$ such that the local rule $f(a_1 a_2 a_3) = b$. For each node N at level i, i $\geq 0$, construct for each a $\in$, a node $N_a$ at level i + 1 as follows. If $a_1 a_2 a_3$ is an element in the set corresponding to N, the set corresponding to $N_a$ will consist of exactly those 3-tuples $a_2 a_3 d$, $d \in \Sigma_2$, for which $f(a_2 a_3 d) = a$. A directed arc labeled a is then drawn from node N to node $N_a$. If for each $a_1 a_2 a_3$ in N there are no such elements d, then this node $N_a$ is not included in the graph and we say that node N is terminal in the graph. We will also say that symbol a made N terminal. If during the construction process a number of nodes appear at the same or different levels, all associated with the same subset of $\Sigma_2^3$, then each will be a distinct node in the graph but only one, arbitrarily chosen, will be extended, i.e., only one will have directed arcs leaving it going to nodes at the next higher level. The equal set nodes not exended will be called frontier nodes. This construction process must eventually terminate since there is a bound on the number of possible subsets of $\Sigma_2^3$.

What this algorithm does is it traces through several iterations of the local rule being applied. Noting where there are terminal nodes is important because this is where we can't generate a specific output, thus the cellular automata is non-surjective. Noting where we have frontier nodes is also important because this identifies where the cellular automata is cyclic. If we can cycle back to every node without running into any terminal nodes, then we can generate all possible outputs, which would confirm surjectivity.

When implementing this algorithm on rule 116 for example, we get a terminal node in level 2 of our graph, proving rule 116 is non-surjective. In the case of this terminal node, its significance is that it shows that rule 116 cannot generate the sequence 010, hence 116 is non-surjective. Here is a trace of the algorithm.

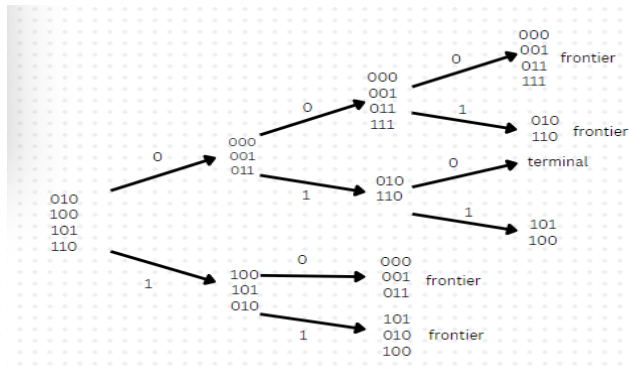

## 5 Determining Surjectivity

Although surjectivity can be determined through the Construction Algorithm, for our purposes we utilized an alternative solution implemented in Mathematica paclet, **KlausSutner/Automata**. The algorithm they implemented to determine surjectivity is more efficient and can also determine whether a CA is open and injective. The **Automata** paclet also provides many other useful functions and algorithms that were useful for our purposes.

### 5.1 Getting Started with Automata.
Using **Automata**, it is extremely easy to represent Cellular Automata:

For Elementary Cellular Automata, you can simply give the rule number (base 10) to the ECA function:

```
c = ECA[90]
```

For general Cellular Automata, you can simply pass the span, size of the alphabet, and rule number (base 10) to the CA function. In fact, ECA is simply just a shortcut for span 3, alphabet $\Sigma_2$ CA.

```
c = CA[4, 2, 3612]

CA[3, 2, 30] == ECA [30]
```

### 5.2 ClassifyCA.
**Automata** provides an extremely useful function called *ClassifyCA* where if given a cellular automata, it will return an enum indicating the one of 3 possibilities:

- 0 → No Special Properties
- 1 → Surjective
- 2 → Open
- 3 → Injective

It should be noted that if a CA is injective, it is also open and surjective and if a CA is open, then it is also surjective.

Before we can analyze how ClassifyCA works, we must first introduce the concepts of Strongly Connected Graphs, Components, and Condensation Graphs [6].

**Strongly Connected Graphs** are defined as graphs in which for all verticies, there exists a path which connects them to every other vertex. It can be determined if a graph is strongly connected in $O(V + E)$ (Linear) time.

**Strongly Connected Components** are the partitions of a graph which are themselves strongly connected. **Trivial** Strongly Connected Components consist of a single vertex which is not connected to itself with an edge. Otherwise, the strongly connected component is **Non-trivial**.

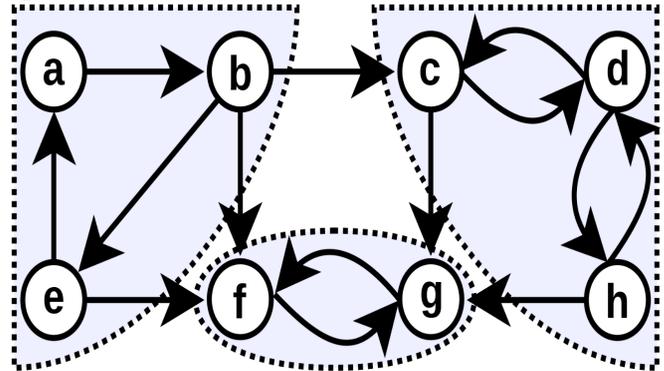

The **Condensation Graph** or **Condensation** of a graph G is a directed, acyclic graph where each vertex represents a set of strongly connected components in G. Note the condensation graph can only be acylic if it contains no strongly connected components with more than 1 vertex.

*ClassifyCA* implements a similar algorithm to the Construction Algorithm in the sense that it turns the Cellular Automata into a Finite State Automata and performs some analysis on the new graph. However, instead of searching for frontier/terminal nodes, *ClassifyCA* checks the properties of the condensation graph of the transition system graph of the Finite State Automata representaton of the Cellular Automata.

Particularly, *ClassifyCA* checks to see if the number of strongly connected components where the state is able to get back to itself is equal to the number of states. If it is, then the cellular automata is surjective.

Additionally, it should be noted that if the number of non-trivial strongly connected components is 1, then the cellular automata is injective. There are other criteria used to determine whether the Cellular Automata is open, but those are beyond the scope of this paper and should be abstracted.

Here is the source code for *ClassifyCA*

```
ClassifyCA[ca_CA,opts:OptionsPattern[]]:=
Module[{n,T,G,scc,ntscc,sccind,ed,H,HH,diag,res},
  If[ !BalancedQCA[ca], Return[0] ];
  n = AlphabetCountCA[ca]^(WidthCA[ca]-1);
  T = TransitionSystem@ToSemiautomatonCA[ca];
```

```
6      G = TransitionSystemGraph[ProductTransitionSystem[T,
           StateType->"Shallow"]];
7      {scc,ntscc,sccind} = StronglyConnectedComponents[G];
8      {H,scc,ntscc} = Most@CondensationGraph[G,Full->True,
           Type->"Shallow"];
9      diag = First@Select[scc,MemberQ[#,{1,1}]&];
10     HH = Subgraph[H,Intersection[
11         FlattenOne[VertexOutComponent[H,#]& /@ ntscc],
12         FlattenOne[VertexOutComponent[ReverseGraph@H,#]& /@
           ntscc]
13         ]];
14
15     res = Which[
16         Length[diag] != n,       0,
17         Length[ntscc]==1,        3,
18         VertexDegree[HH,diag]==0, 2,
19         True,                    1 ];
20     If[ OptionValue[Full], Sow[{res,H,HH,scc,ntscc}] ];
21     res
22   ];
```

### 5.3 Application.

We can utilize the *ClassifyCA* and the Table function to be able to apply and record which CA rules are surjective. For example, this code finds all of the surjective CA of width 4 between Rule 0 and Rule 10,000.

```
1 Cases[Table[{r, ClassifyCA[CA[4,2,r]]},{r, 0, 10000}],{x_
      ,y_} /; $y>0$]
```

Another interesting application that we can use is given a local tabular rule in polynomial form, we can determine what CA is associated with this function and then determine if it is surjective. For example, this code snippet determines whether the CA associated with the function $f(x_0,x_1,x_2,x_3,x_4,x_5) = (x_0 + x_3 + x_2 * x_5$ mod 2) is surjective is:

```
1 c = ConvertToCA[{a_,b_,c_,d_,e_,f_}:->Mod[a+d+c*f,2],6,2,
      Type->Rule]
2 ClassifyCA[c]
```

## 6  FDense

Bryant Lee created three programs that were used to study one-dimensional CA. The first was FDense which is an algorithm used to determine (given a cellular automata *f*, an integer N ≥ 2, a positive integer m, and a finite set *K* of positive integer) whether the set $Per(f) \cap P_k(S_N)$ is m-dense. The limitations on his program, with N=2, was m = 13 and k = 27 since the algorithm demanded a lot of memory space. For this reason, their data set was restricted entirely to the case of N = 2 symbols. Bryant created his program in 2006 using c++ which can no longer be compiled using the current version of g++. Therefore, to recreate his program, we created a pseudo-code for FDense.

```
1  pseudo-code:
2    Define a function FDense that takes as input a c.a. f,
3    an integer N, a positive integer m, and a set of
4    positive integers K.
5      For each value of k in K:
6        Compute the set Per(f) /\ Pk(SN) of jointly
7        periodic points of f with period k.
8        If this set is empty, f is not periodic with period
           k.
9        Otherwise, for each word w of length m on the
10       alphabet A:
11         Check whether there exists a point x in
12         $Per(f) /\ Pk(SN)$ such that w appears as a
13         substring of x.
14         If no such point x exists, f is not m-dense at k.
15     If all words of length m on the alphabet A appear as
16     substrings in some point of $Per(f) /\ Pk(SN)$, f is
17     m-dense at k.
```

To test our program, we are first going to create the program for span 4 cellular automata. The way we are going to implement the code is to make it so that every word of length m is periodic under the shift k where k ≥ m. Then, using the tabular rules defined in Table 1 of [1], we are going to see if the words are also periodic under the c.a. If the words are periodic under c.a the, by defintion, they are jointly periodic points. If every word of length m exists in a sequence in the set $Per(f) \cap Pk(SN)$ is jointly periodic, then the c.a. map is m dense.

We first generates all words of length m which gives a total of $2^m$. In order to make the words periodic for k m, we generate all words of length (k-m) which gives a total of $2^{k-m}$. Then we concatenated all words of length (k-m) to every word of length m which gives us a total of $2^k$ words.

Similar to [1], we will also use the surjective c.a. of span 4 and span 5 from [13]. The first table is from [1, Table 1] which contains 32 span 4 surjective c.a. The second table from [1, Table 2] contains the 26 span 5 surjective maps. We will compare the results of our programs to the results of [1, Table 3]. Our source code below is implemented in C.

```c
1  #include <stdio.h>
2  #include <stdlib.h>
3  #include <math.h>
4  #include <string.h>
5
6  #define N 2
7
8  typedef struct Rule {
9      int key;
10     int value;
11 } Rule;
12
13 typedef struct StringArray {
14     char **strings;
15     int size;
16 } StringArray;
17
18 Rule rule[16];
19
20 void createHashMap(const char *tabularRule, int k) {
21     int n = (int) pow(2, k);
22
23     for (int i = 0; i < n; i++) {
24         rule[i].key = i;
25         rule[i].value = tabularRule[i] - '0';
26     }
27 }
28
29 void generateWordsHelper(const char *prefix, int length,
       StringArray *words) {
30     if (strlen(prefix) == length) {
31         words->strings[words->size] = strdup(prefix);
32         words->size++;
33     } else {
34         for (int i = 0; i < N; i++) {
35             char newPrefix[strlen(prefix) + 2];
36             strcpy(newPrefix, prefix);
37             newPrefix[strlen(prefix)] = '0' + i;
38             newPrefix[strlen(prefix) + 1] = '\0';
39             generateWordsHelper(newPrefix, length, words)
              ;
40         }
41     }
42 }
43
44 StringArray generateWords(int length) {
45     int max_size = (int) pow(N, length);
46     StringArray words;
47     words.strings = (char **) malloc(max_size * sizeof(
          char *));
48     words.size = 0;
49     generateWordsHelper("", length, &words);
50     return words;
51 }
52
53 char *applyRule(const char *word, int span) {
54     int key;
55     char *newWord = (char *) malloc((strlen(word) + 1) *
          sizeof(char));
56     newWord[strlen(word)] = '\0';
57
58     for (int i = 0; i < strlen(word); i++) {
59         char pattern[span + 1];
60         pattern[span] = '\0';
61
62         for (int j = 0; j < span; j++) {
```

```
                pattern[j] = word[(i + j) % strlen(word)];
            }

        key = (int) strtol(pattern, NULL, 2);

        for (int j = 0; j < 16; j++) {
            if (rule[j].key == key) {
                newWord[i] = '0' + rule[j].value;
                break;
            }
        }
    }

    return newWord;
}

int main() {
    int k = 20;
    int m = 10;
    int span = 4;

    StringArray originalWords = generateWords(m);
    StringArray wordsWithKperiod = generateWords(k - m);
    StringArray allWords;
    allWords.strings = (char **) malloc(originalWords.
     size * wordsWithKperiod.size * sizeof(char *));
    allWords.size = 0;

    for (int i = 0; i < originalWords.size; i++) {
        for (int j = 0; j < wordsWithKperiod.size; j++) {
            char *newWord = (char *) malloc((strlen(
     originalWords.strings[i]) + strlen(wordsWithKperiod.
     strings[j]) + 1) *sizeof(char));
            strcpy(newWord, originalWords.strings[i]);
            strcat(newWord, wordsWithKperiod.strings[j]);
            allWords.strings[allWords.size] = newWord;
            allWords.size++;
        }
    }

    int true = 0;
    for (int i = 0; i < allWords.size; i++){
        char *word = allWords.strings[i];
        char *newWord = applyRule(word, span);

        while (strcmp(newWord, word) != 0) {
            char *temp = newWord;
            newWord = applyRule(newWord, span);
            free(temp);
        }
        true += 1;
        free(newWord);
    }

    for (int i = 0; i < originalWords.size; i++) {
        free(originalWords.strings[i]);
    }
    for (int i = 0; i < wordsWithKperiod.size; i++) {
        free(wordsWithKperiod.strings[i]);
    }
    for (int i = 0; i < allWords.size; i++) {
        free(allWords.strings[i]);
    }

    free(originalWords.strings);
    free(wordsWithKperiod.strings);
    free(allWords.strings);

    printf("\n");
    return 0;
}
```

**6.1 Code analysis.** To analyze the algorithmic complexity of the code, let's break it down into the major functions:

**generateWordsHelper:** This is a recursive function that generates all possible binary words of a given length. The complexity is $O(2^L)$, where L is the length of the words generated. This is because each position in the word has 2 possibilities (0 or 1), and there are L positions.

**generateWords:** This function is a wrapper for generateWordsHelper and has the same complexity, $O(2^L)$.

**applyRule:** This function iterates through each character in a word and applies a rule based on a span. The complexity is $O(S * W)$, where S is the size of the span and W is the length of the word.

**main:** Generating originalWords and wordsWithKperiod takes $O(2^m)$ and $O(2^{k-m})$ time complexity, respectively. Creating allWords by concatenating each pair of words from originalWords and wordsWithKperiod takes $O(2^k)$ time complexity, as there are $2^k$ combined words. Applying the rule and comparing words in a while loop takes $O(2^k * S * W)$ time complexity, where S is the span size and W is the length of the words in allWords.

Overall, the time complexity of the code is $O(2^k * S * W)$. Note that $2^k$ grows exponentially with k, making the algorithm's performance degrade rapidly as k increases. The statement describes a limitation of the algorithm when dealing with large values of k, as it consumes a significant amount of memory. The main reason for this memory consumption is the generation and storage of all possible binary words for a given k. When we generate allWords in the main function, we are essentially creating $2^k$ binary strings. Each string is stored in memory, and as k increases, the number of strings grows exponentially. For instance: k = 23: There are $2^23$ = 8,388,608 binary strings. k = 26: There are $2^26$ = 67,108,864 binary strings. As you can see, the number of binary strings generated grows rapidly with k, leading to high memory consumption. For large values of k, storing all the binary strings in memory may not be possible as $k \to \infty$, as it can exhaust the available memory resources. Our algorithm seem to face the same limitation that Mike Boyle and Bryant Lee faced. However, we are able to run our program on CoCalc which provides us with memory exclusive to our local machine. Even then, the time that it takes to iterate through every word still takes a very long time when k is larger than 24.

## 7 Results

| Map | 10-dense | 13-dense at | Map | 10-dense | 13-dense |
|---|---|---|---|---|---|
| 1 | 11,13,15,17 | 13,15,17 | 17 | [] | [] |
| 2 | 10-20 | 13-21 | 18 | [] | [] |
| 3 | 18 | 18 | 19 | [] | [] |
| 4 | [] | [] | 20 | [] | [] |
| 5 | [] | [] | 21 | [] | [] |
| 6 | 10-21 | 13-19 | 22 | [] | [] |
| 7 | 10-18 | 13-18 | 23 | [] | [] |
| 8 | [] | [] | 24 | [] | [] |
| 9 | 11,13,15,17,19 | 13,15,17,19 | 25 | [] | [] |
| 10 | [] | [] | 26 | 11-20 | 13-18 |
| 11 | [] | [] | 27 | [] | [] |
| 12 | [] | [] | 28 | [] | [] |
| 13 | 11,13,15,17 | 13,15 | 29 | [] | [] |
| 14 | [] | [] | 30 | [] | [] |
| 15 | [] | [] | 31 | [] | [] |
| 16 | 10-18 | 13-18 | 32 | [] | [] |

**Table 1** This are the results of our FDENSE program. Using the 32 span 4 onto maps of [1, Table 1], we used our program to see if the maps are 10-dense and 13-dense for values of 24 ≥ k ≥ m. [] means that we were not able to get any conclusive result for that map so far/not tested yet.

| Map | Tabular Rule | Map | Tabular Rule |
|---|---|---|---|
| 1 | 0000 1111 0010 1101 | 17 | 0011 1001 1100 1100 |
| 2 | 0011 1001 1100 1100 | 18 | 0011 1010 0011 1100 |
| 3 | 0001 1100 0011 1110 | 19 | 0001 1100 0011 1110 |
| 4 | 0001 1110 0101 1010 | 20 | 0011 1100 0101 0011 |
| 5 | 0010 1001 0110 1101 | 21 | 0011 1100 0101 1100 |
| 6 | 0010 1101 0000 1111 | 22 | 0011 1100 1010 0011 |
| 7 | 0011 0011 0110 0011 | 23 | 0011 1100 1010 1100 |
| 8 | 0011 0011 0110 1100 | 24 | 0011 1110 0001 1100 |
| 9 | 0011 0011 1001 0011 | 25 | 0100 1001 0110 1011 |
| 10 | 0011 0011 1001 1100 | 26 | 0100 1011 0000 1111 |
| 11 | 0011 0101 0011 1100 | 27 | 0101 1010 0001 1110 |
| 12 | 0011 0101 1100 0011 | 28 | 0101 1010 0111 1000 |
| 13 | 0011 0110 0011 0011 | 29 | 0110 1011 0100 1001 |
| 14 | 0011 0110 1100 1100 | 30 | 0110 1101 0010 1001 |
| 15 | 0011 1000 0111 1100 | 31 | 0111 1000 0101 1010 |
| 16 | 0011 1001 0011 0011 | 32 | 0111 1100 0011 1000 |

**Table 2** This is a table of the list of span 4 surjective cellular automata that were used in [1]. We wanted to see if our program is correct so we used the same set of inputs and compared our output to [1, Table 3]. The maps described in this table is taken from [7, table I].

## 8 Conclusion

Mike Boyle proposed that every span 4 surjective cellular automata is at least 13 dense [1]. While we were replicating his experiment for the maps described in table 1, our results seem to match his proposal. We wanted to extend his proposition to span 6 surjective cellular automata. However, we were not able to quantify his proposal for majority of the surjective maps that we found because our FDense program took too long to run and it is inconclusive for big values of m and k.

## 9 Acknowledgement

We thank you Professor Rodrigo Trevino and the Lab of Experimental Mathematics for giving us the opportunity to work with graduate student, Chi-Hao. We were able to learn a lot about computational theory and also had the opportunity to present our work to our peers.

## Appendix A: List of Surjective Span 6 CA Rules

```
1  11913610175290432170,
2  12273810184465984170,
3  10778685752149174890,
4  9838244740979984520,
5  9833478358074685320,
6  9765923332887705720,
7  13527612318716019780,
8  15987178194065302050,
9  12297829015969158570,
10 6149008516228754090,
11 8608480568017455240,
12 9761684715563616120,
13 11067990149230388838,
14 12273810549538182570,
15 11936045731524794970,
16 6531808643310593370,
17 18446744069414584320,
18 18446462603027742720,
19 18374966859414961920,
20 17361641481138401520,
21 14757395258967641292,
22 12297829382473034410,
23 12297735558912453290,
24 12273904009452628650,
25 11913616039262988970,
26 10760694534656141994,
27 12273810549532633770,
28 12273810550964267690,
29 12297735922558610090,
30 12297735923984673450,
31 11937535914725255850,
32 11914929955658181290,
33 11053616526923573930,
34 11072831592130587306,
35 11915017572991019690,
36 12273903644380408490,
37 12297829381046949290,
38 11915017571643578970,
39 11053691000511161958,
40 11937535913383058090,
41 11072831590989719210,
42 10766582143467022954,
43 11936128518366866090,
44 11936045731440601770,
45 11915017571564934570,
46 12297829381125593690,
47 10851025924903036518,
48 11936045732872235690,
49 10851025926048361130,
50 12297741076598008490,
51 10850959696507349674,
52 12275311039396600490,
53 10834137168606815914,
54 11140386616255343210,
55 11068046444512062122,
56 11067990148944079530,
57 11053691000230401450,
58 10851025924700920410,
59 12297829381327709798,
60 11067990150375713450,
61 11053691001656486570,
62 11053634925137406634,
63 10850972941984377514,
64 12297754322479262378,
65 10837514919665621674,
66 12278688790858065578,
67 12008468690110147178,
68 6149102339789291520,
69 6196766168842283520,
70 6872316420712079520,
71 12297923204601937920,
72 12273717456122085120,
73 11935962946030203120,
74 11067933855094066380,
75 12249885541595589120,
76 12321755119128433920,
77 11893906626277563120,
78 11039335557662271180,
79 18398892593240473770,
80 18446556422293465770,
81 11574355905568809120,
82 11556131500342665120,
83 12659530245063331920,
84 10634005406540393580,
85 17723342341370677770,
86 18446593951717689480
```

# List of Figures

# List of Tables